\documentclass[aps,12pt,pra,showpacs]{revtex4}
\usepackage{amsmath}
\usepackage{color}
\usepackage{graphicx}
\usepackage{dcolumn}
\usepackage{bm}

\begin{document}

\title{Realization of Topological Quantum Computation with surface codes}
\author{Su-Peng Kou}
\affiliation{Department of Physics, Beijing Normal University, Beijing 100875, China}

\begin{abstract}
In this paper, the degenerate ground states of $Z_{2}$ topological order on
a plane with holes (the so-called surface codes) are used as the protected
code subspace to build a topological quantum computer by tuning their
quantum tunneling effect. Using a designer Hamiltonian - the Kitaev
toric-code model as an example, we study quantum tunneling effects of the
surface codes and obtain its effective theory. Finally, we show how to do
topological quantum computation including the initialization, the unitary
transformation and the measurement.
\end{abstract}

\pacs{03.67.Lx, 03.75.Lm, 75.45.+j, 75.10.Jm} \maketitle

\section{Introduction}

Quantum computers are predicted to utilize quantum states to process tasks
far faster than those of conventional classical computers. Various designs
have been proposed to build a quantum computer, such as manipulating
electrons in a quantum dot or phonon in ion traps, cavity QED, nuclear spin
by NMR techniques. DiVincenzo pointed out that in order to build a real
quantum computer\cite{de}, the following five criterions should be satisfied
: 1) Scalability of extendible qubits; 2) Initialization (Creation of highly
entangled states); 3) Local operations on multi-qubit; 4) Measurement of
entangled states; 5) Long decoherence time. In particular, the fifth
criteria (low decoherence condition) becomes a trouble to reach the goal.

{Recently, people find that it may be possible to incorporate intrinsic
fault tolerance into a quantum computer - topological quantum computation
(TQC) which} has the debilitating effects of decoherence and free from
errors. The key point is to store and manipulate quantum information in a
\textquotedblleft non-local\textquotedblright \ way, namely, the
\textquotedblleft non-local\textquotedblright \ properties of a quantum
system remain unchanged under local operations. An interesting idea to
realize fault-tolerant quantum computation is anyon-braiding, proposed by
Kitaev\cite{k1,k2}. He pointed out that the degenerate ground states of a
topological order make up a protected code subspace (the topological qubit)
free from error\cite{wen,ioffe}.

Topological order is a new type of quantum orders beyond Landau's symmetry
breaking paradigm\cite{wen1,wen2,wen3}. People know that there are two types
of topological orders in two dimensional $S=1/2$ spin models - non-Abelian
topological ordered state and $Z_{2}$ topological ordered state. Those
topological ordered states may appear in frustrated spin systems or dimer
models\cite{spin1,spin2,spin3}. Because of full gapped excitations, these
topological orders are robust against any perturbations, even those
perturbations that break all symmetries.

In non-Abelian topological orders, the elementary excitation are non-Abelian
anyon with nontrivial statistics. Now people focus on realizing TQC by
braiding non-Abelian anyons\cite{k2,sa1}. The degenerate states undergoes a
nontrivial unitary transformation when a non-Abelian anyon moves around the
other. One can initial, manipulate and measure the degenerate ground states
with several non-Abelian anyons\cite{k2,sa1}, which has become a hot issue
recently \cite{pa,sarma,du1,zoller,vid,vids,zhang,zhang1,gao,kou1}.

On the contrary, $Z_{2}$ topological order is the simplest
topological ordered state with three types of quasi-particles:
$Z_{2}$ charge, $Z_{2}$ vortex, and fermions\cite{wen4}. $Z_{2}$
charge and $Z_{2}$ vortex are all bosons with mutual $\pi $
statistics between them. The fermions can be regarded as bound
states of a $Z_{2}$ charge and a $Z_{2}$ vortex. In the last decade,
several exactly solvable spin models with $Z_{2}$ topological orders
were found, such as the Kitaev toric-code model \cite{k1}, the Wen's
plaquette model \cite{wen4,wen5} and the Kitaev model on a honeycomb
lattice \cite{k2}. In Ref.\cite{kou1}, an alternative way to design
TQC is proposed by manipulating the toric codes - the protected code
subspace of $Z_{2}$ topological orders. To manipulate the degenerate
ground states we can tune tunneling by controlling external field on
spin models. {However, }to build a real quantum computer,
multi-qubit is necessary. It is indeed a challenge to realize a spin
model on a manifold with higher genus in experiments. To solve this
problem, in this paper we design TQC by manipulating the surface
code (See detail in the main content) rather than toric code, since
it is more easy to make a hole in a surface.

The paper is organized as follows. In Sec. II, we study the properties of
the surface code. In Sec. III, an effective theory of the surface code in $%
Z_{2}$ topological orders is formalized. By using the Kitaev toric-code
model as an example, we demonstrate how to control the surface code by
tuning the tunneling of the degenerate ground states\cite{wen,wen1,wen2,wen3}%
. In Sec. IV, based on the effective theory, the TQC is shown including the
initialization, the unitary transformation and the measurement. Finally, we
draw the conclusions in Sec. V.

\section{Degenerate ground states as surface codes}

In the paper, we demonstrate the TQC in $Z_{2}$ topological order\emph{\ }by
using a designed model - the Kitaev toric-code model (an effective model of
the Kitaev model on two dimensional hexagonal lattice) as an example\cite%
{k1,k2,wen}. Here the Hamiltonian of the Kitaev toric-code model is
described by \cite{k1}
\begin{equation}
H=-g(\sum \limits_{i\in \mathrm{even}}{Z_{i}}+\sum \limits_{i\in \mathrm{odd}%
}{X_{i})}.
\end{equation}%
where
\begin{equation*}
{Z_{i}=}\sigma _{i}^{z}\sigma _{i+\hat{e}_{x}}^{z}\sigma _{i+\hat{e}_{x}+%
\hat{e}_{y}}^{z}\sigma _{i+\hat{e}_{y}}^{z},\text{ }{X_{i}=}\sigma
_{i}^{x}\sigma _{i+\hat{e}_{x}}^{x}\sigma _{i+\hat{e}_{x}+\hat{e}%
_{y}}^{x}\sigma _{i+\hat{e}_{y}}^{x}
\end{equation*}%
with $g>0.$ $\sigma _{i}^{x,y,z}$ are Pauli matrices on sites $i.$\ See the
scheme in Fig.1.

\begin{figure}[tbp]
\includegraphics[clip,width=0.3\textwidth]{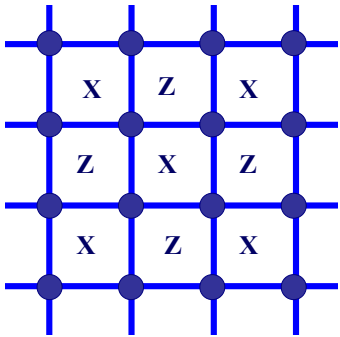}
\caption{The scheme of the Kitaev toric-code model }
\label{Fig.1}
\end{figure}
Firstly we study the ground state degeneracy $\mathcal{Q}$ of the Kitaev
toric-code model. The ground state is a $Z_{2}$ topological state that is
denoted by ${Z_{i}=X_{i}\equiv +1}$ at each site with energy,
\begin{equation}
E_{0}=-g\mathcal{N}
\end{equation}%
where $\mathcal{N}$ is the total lattice number\cite{wen,wen4,wen5,kou2}.
Under the periodic boundary condition (on a torus), the degeneracy $\mathcal{%
Q}$ is dependent on $\mathcal{N}$ : $\mathcal{Q}=4$ on even-by-even ($e\ast
e $) lattice, $\mathcal{Q}=2$ on other cases (even-by-odd ($e\ast o$),
odd-by-even ($o\ast e$) and odd-by-odd ($o\ast o$) lattices)\cite%
{wen,wen4,wen5,kou2}. In addition, on a manifold with high genus ($\chi >1${%
), }$\mathcal{Q}${\ becomes }$4^{\chi }${\ }on $e\ast e$ lattice and{\ }$%
4^{\chi }-2${\ }on other cases. Kitaev have noted the degenerate
ground states of the Kitaev toric-code model on a torus as \emph{the
toric code}. On the other hand, for the ground states of the Kitaev
toric-code model {o}n a surface with open boundary condition, the
ground state degeneracy now becomes
\begin{equation}
\mathcal{Q}=2^{n}
\end{equation}%
(without considering the edge states)\cite{hole,edge}. Here $n$ is the
number of holes. In the following part we call the degenerate ground states
of the Kitaev toric-code model {o}n a surface with open boundary condition
\emph{the surface codes}.

To classify the degeneracy of the ground states (the surface codes), we
define three types of closed string operators $W_{c}(\mathcal{C}),$ $W_{v}(%
\mathcal{C})$ and $W_{f}(\mathcal{C})=W_{c}(\mathcal{C})W_{v}(\mathcal{C}).$
Here $W_{c}(\mathcal{C})=\prod_{\mathcal{C}}\sigma _{i}^{x}$ (or $W_{v}(%
\mathcal{C})=\prod_{\mathcal{C}}\sigma _{i}^{z}$) is the products of spin
operators along a loop $C$ connecting even-plaquettes (or odd-plaquettes) of
neighboring links, with $\mathcal{C}$ denoting closed loops. It is obvious
that the three types of closed string operators ($W_{c}(\mathcal{C}),$ $%
W_{v}(\mathcal{C})$, $W_{f}(\mathcal{C})$) correspond to three
quasi-particles ($Z_{2}$ charge, $Z_{2}$ vortex, and fermions) respectively.
One can easily check the commutation relations between the closed string
operators and the Hamiltonian
\begin{equation}
\left[ H,\text{ }W_{c}(\mathcal{C})\right] =\left[ H,\text{ }W_{v}(\mathcal{C%
})\right] =\left[ H,\text{ }W_{f}(\mathcal{C})\right] =0.
\end{equation}

\begin{figure}[tbp]
\includegraphics[clip,width=0.5\textwidth]{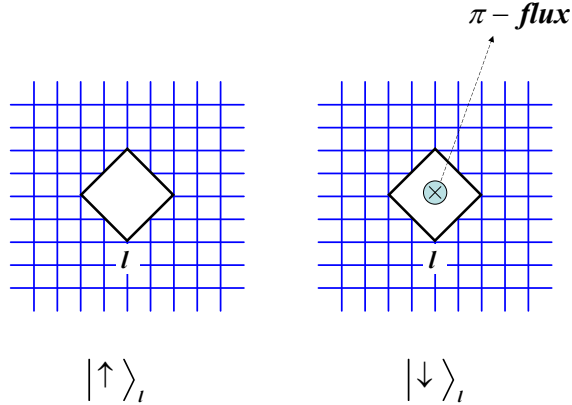}
\caption{Mapping the qubit of a hole onto a pseudo-spin}
\label{Fig.2}
\end{figure}

In particular, for the surface codes, we can define two types of special
closed string operators, $W_{v}(\mathcal{C}_{A})$ and $W_{f}(\mathcal{C}%
_{B}).$ Here $\mathcal{C}_{A}$ denotes a closed loop around a hole (labeled
by an index $l$) and $\mathcal{C}_{B}$ denotes a loop from the hole to the
boundary of the system. Although $W_{f}(\mathcal{C}_{B})$ is not the
original closed string operator, its topological properties are the same.
Due to the anti-commutation relation between $W_{v}(\mathcal{C}_{A})$ and $%
W_{f}(\mathcal{C}_{B})$,
\begin{equation}
\left \{ W_{v}(\mathcal{C}_{A}),W_{f}(\mathcal{C}_{B})\right \} =0,
\end{equation}%
\ we may identify $W_{v}(\mathcal{C}_{A})$ and $W_{f}(\mathcal{C}_{B})$ as
pseudo-spin ($S=\frac{1}{2}$)\ operators $\tau _{l}^{z}$ and $\tau _{l}^{x}$%
, respectively. The ground states become the eigenstates of $\tau _{l}^{z}$.
Then one has two degenerate ground states (denoted by $\mid m_{l}\rangle $)
for the case with a single hole. For $m_{l}=0,$ we have
\begin{equation}
\tau _{l}^{z}\mid m_{l}\rangle =\mid m_{l}\rangle ,
\end{equation}%
and for $m_{l}=1$ we have
\begin{equation}
\tau _{l}^{z}\mid m_{l}\rangle =-\mid m_{l}\rangle .
\end{equation}%
Physically, the topological degeneracy arises from presence or the absence
of $\pi $ flux of $Z_{2}$ vortex through the hole (See Fig.2). The values of
$m_{l}$ reflect the presence ($m_{l}=1$) or the absence ($m_{l}=0$) of the $%
\pi $ flux in the hole.

\begin{figure}[tbp]
\includegraphics[clip,width=0.5\textwidth]{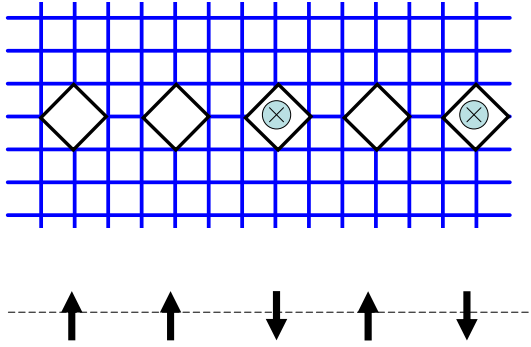}
\caption{Mapping the qubits of a line of holes onto a pseudo-spin chain}
\label{Fig.3}
\end{figure}

Furthermore for the $Z_{2}$ topological order with $n$ holes, the degenerate
ground states have $2^{n}$ fold degeneracy as Eq.(3). The $2^{n}$ degenerate
ground states can be mapped onto a $2^{n}$-level quantum system of
pseudo-spins $\mathbf{\hat{\tau}}_{l}$ by the following correspondence,
\begin{equation*}
\mid m_{l}=0\rangle \rightarrow \mid \uparrow \rangle _{l},\text{ }\mid
m_{l}=1\rangle \rightarrow \mid \downarrow \rangle _{l}.
\end{equation*}%
We now get $2^{n}$ degenerate ground states of a chain of $n$ holes. The $%
2^{n}$ degenerate ground states are denoted by
\begin{equation*}
\mid m_{1},...,m_{l}...,m_{n}\rangle =\mid m_{1}\rangle \otimes ...\otimes
\mid m_{l}\rangle ...\otimes \mid m_{n}\rangle ,
\end{equation*}%
where the two degenerate ground states of each hole are denoted by $\mid
m_{l}\rangle ,$ $m_{l}=0,$ $1$. In the following parts we use the degenerate
ground states on a surface with a chain of holes to do TQC (See Fig.3).

\section{Perturbative approach of quasi-particles}

In this section we study the properties of the quasi-particles. In this
solvable model, $Z_{2}$ vortex is defined as ${Z_{i}=-1}$ at even
sub-plaquette and $Z_{2}$ charge is ${X_{i}=-1}$ at odd sub-plaquette. The
mass gap of $Z_{2}$ charge and $Z_{2}$ vortex is $2g$. In particular, the
bound states of a $Z_{2}$ charge and a $Z_{2}$ vortex on two neighbor
plaquettes obey fermionic statistic. All quasi-particles in this model have
flat bands. The energy spectra are $E_{v}=E_{c}=2g$ for $Z_{2}$ vortex and $%
Z_{2}$ charge, $E_{f}=4g$ for fermions, respectively. In other words, the
quasi-particles cannot move at all.

Under the perturbation
\begin{equation}
\hat{H}^{\prime }=\sum \limits_{i}\mathbf{h}\cdot \mathbf{\sigma }%
_{i}=h^{x}\sum \limits_{i}\sigma _{i}^{x}+h^{y}\sum \limits_{i}\sigma
_{i}^{y}+h^{z}\sum \limits_{i}\sigma _{i}^{z},
\end{equation}%
the quasi-particles begin to hop. The term $h^{x}\sum \limits_{i}\sigma
_{i}^{x}$ drives the $Z_{2}$ vortex without affecting fermion and $Z_{2}$
charge. For a $Z_{2}$ vortex at $i$ plaquette ${X_{i}=-1,}$ when $\sigma
_{i}^{x}$ acts on $i+\hat{e}_{x}$ site, it hops to $i+\hat{e}_{x}-\hat{e}%
_{y} $ plaquette denoted by ${X_{i+\hat{e}_{x}-\hat{e}_{y}}=-1},$%
\begin{equation}
{X_{i}}{=+1\rightarrow X_{i}=-1,}\text{ }{X_{i+\hat{e}_{x}-\hat{e}_{y}}}{%
=+1\rightarrow X_{i+\hat{e}_{x}-\hat{e}_{y}}=-1.}  \notag
\end{equation}
Moreover, a pair of $Z_{2}$ vortices at $i$ and $i+\hat{e}_{x}-\hat{e}_{y}$
plaquettes can be created by the operation of $\sigma _{i}^{x},$%
\begin{equation}
{X_{i}}{=+1\rightarrow X_{i}=-1,}\text{ }{X_{i+\hat{e}_{x}-\hat{e}_{y}}}{%
=+1\rightarrow X_{i+\hat{e}_{x}-\hat{e}_{y}}=-1}\text{.}  \notag
\end{equation}
Similarly, the term $h^{z}\sum \limits_{i}\sigma _{i}^{z}$ drives the $Z_{2}$
charge without affecting fermion and $Z_{2}$ vortex. In particular, there
exist two types of fermions : the fermions on the vertical links and the
fermions on the parallel links. The term $h^{y}\sum \limits_{i}\sigma
_{i}^{y} $ drives fermions hopping without affecting $Z_{2}$ vortex and $%
Z_{2}$ charge : the fermions on the vertical links move vertically and the
fermions on the parallel links move parallelly. That means both types of
fermions cannot turn round any more.

To describe the dynamics of the quasi-particles we use the perturbative
approach in Ref.\cite{zoller,vid,vids,vid1,vid2,kou1}. In the perturbative
approach, the spin operators are represented by hopping terms of
quasi-particles,%
\begin{eqnarray}
\sigma _{i}^{x} &\rightarrow &(\phi _{1,i}^{\dag }\phi _{1,i+e_{x}\pm
e_{y}}+\phi _{1,i}^{\dag }\phi _{1,i+e_{x}\pm e_{y}}^{\dag }+h.c.), \\
\sigma _{i}^{z} &\rightarrow &(\phi _{2,i}^{\dag }\phi _{2,i+e_{x}\pm
e_{y}}+\phi _{2,i}^{\dag }\phi _{2,i+e_{x}\pm e_{y}}^{\dag }+h.c.),  \notag
\\
\sigma _{i}^{y} &\rightarrow &(\phi _{3,i_{3}}^{\dag }\phi _{3,i_{3}\pm
e_{x}}+\phi _{3,i_{3}}^{\dag }\phi _{3,i_{3}\pm e_{x}}^{\dag }+h.c.)  \notag
\\
&&+(\phi _{4,i_{4}}^{\dag }\phi _{4,i_{4}\pm e_{y}}+\phi _{4,i_{4}}^{\dag
}\phi _{4,i_{4}\pm e_{y}}^{\dag }+h.c.).  \notag
\end{eqnarray}%
Here $\phi _{\alpha ,i_{\alpha }}^{\dag }$ ( $\alpha =1,2$ ) are the
generation operator of $Z_{2}$ vortex, $Z_{2}$ charge and $\phi _{\alpha
,i_{\alpha }}^{\dagger }$ ( $\alpha =3,4$ ) are the generation operator of
fermions, respectively. $i_{1}$ denotes the position on even sub-plaquette
and $i_{2}$ denotes the position on odd sub-plaquette. $i_{3}$ denotes the
position on the vertical links and $i_{4}$ denotes the position on the
parallel links. In addition, one should add a single occupation constraint
(hard-core constraint) as
\begin{equation}
(\phi _{\alpha ,i}^{\dag })^{2}|\Psi \rangle =|\Psi \rangle \text{ or }\phi
_{\alpha ,i}^{\dag }|\Psi \rangle =\phi _{\alpha ,i}|\Psi \rangle
\end{equation}%
where $|\Psi \rangle $ denotes quantum state of the Kitaev toric-code model.
Therefore, by the perturbation method, the Hamiltonian $\hat{H}=\hat{H}_{0}+%
\hat{H}^{\prime }$ can be represented by generation (or annihilation)
operators of quasi-particles,%
\begin{eqnarray}
\hat{H}_{0} &\rightarrow &2g\sum \limits_{i_{1}}\phi _{1,i_{1}}^{\dag }\phi
_{1,i_{1}}+2g\sum \limits_{i_{2}}\phi _{2,i_{2}}^{\dag }\phi _{2,i_{2}}
\notag \\
&&+4g\sum \limits_{i_{3}}\phi _{3,i_{3}}^{\dag }\phi _{3,i_{3}}+4g\sum
\limits_{i_{4}}\phi _{4,i_{4}}^{\dag }\phi _{4,i_{4}}
\end{eqnarray}%
and
\begin{eqnarray}
\hat{H}^{\prime } &\rightarrow &-h^{x}\sum \limits_{\langle
i_{1},j_{1}\rangle }(\phi _{1,i_{1}}^{\dag }\phi _{1,j_{1}}+\phi
_{1,i_{1}}^{\dag }\phi _{1,j_{1}}^{\dagger }) \\
&&-h^{z}\sum \limits_{\langle i_{1},j_{1}\rangle }(\phi _{2,i_{2}}^{\dag
}\phi _{2,j_{2}}+\phi _{2,i_{2}}^{\dag }\phi _{2,j_{2}}^{\dagger })  \notag
\\
&&-h^{y}\sum_{i_{3}}(\phi _{3,i_{3}}^{\dag }\phi _{3,i_{3}\pm e_{x}}+\phi
_{3,i_{3}}^{\dag }\phi _{3,i_{3}\pm e_{x}}^{\dag })  \notag \\
&&-h^{y}\sum_{i_{4}}(\phi _{4,i_{4}}^{\dag }\phi _{4,i_{4}\pm e_{y}}+\phi
_{4,i_{4}}^{\dag }\phi _{4,i_{4}\pm e_{y}}^{\dag })  \notag \\
&&-h^{x}h^{z}\sum_{\left \langle i_{3},j_{4}\right \rangle }(\phi
_{3,i_{3}}^{\dag }\phi _{4,j_{4}}+\phi _{3,i_{3}}^{\dag }\phi
_{4,j_{4}}^{\dag })+h.c..  \notag
\end{eqnarray}

In the following parts, we consider only the perturbation as $\hat{H}%
^{\prime }=h^{x}\sum \limits_{i}\sigma _{i}^{x}+h^{y}\sum \limits_{i}\sigma
_{i}^{y}.$ The perturbative Hamiltonian of the quasi-particles becomes $\hat{%
H}_{v}+\hat{H}_{c}+\hat{H}_{f}$ where
\begin{widetext}
\begin{eqnarray}
\hat{H}_{v} &=&2g\sum \limits_{i_{1}}\phi _{1,i_{1}}^{\dag }\phi
_{1,i_{1}}-h^{x}\sum \limits_{\langle i_{1},j_{1}\rangle }(\phi
_{1,i_{1}}^{\dag }\phi _{1,j_{1}}+\phi _{1,i_{1}}^{\dag }\phi
_{1,j_{1}}^{\dagger })+h.c.  \notag \\
\hat{H}_{c} &=&2g\sum \limits_{i_{2}}\phi _{2,i_{2}}^{\dag }\phi _{2,i_{2}}
\notag \\
\hat{H}_{f} &=&4g\sum \limits_{i_{3}}\phi _{3,i_{3}}^{\dag }\phi
_{3,i_{3}}+4g\sum \limits_{i_{4}}\phi _{4,i_{4}}^{\dag }\phi
_{4,i_{4}}-h^{y}\sum_{i_{3}}(\phi _{3,i_{3}}^{\dag }\phi _{3,i_{3}\pm
e_{x}}+\phi _{3,i_{3}}^{\dag }\phi _{3,i_{3}\pm e_{x}}^{\dag })  \notag \\
&&-h^{y}\sum_{i_{4}}(\phi _{4,i_{4}}^{\dag }\phi _{4,i_{4}\pm e_{y}}+\phi
_{4,i_{4}}^{\dag }\phi _{4,i_{4}\pm e_{y}}^{\dag })+h.c..
\end{eqnarray}
\end{widetext} That means $Z_{2}$ charge cannot move any more.

For the perturbative Hamiltonian on $L_{y}\times L_{x}$ square lattice ($%
L_{y},$ $L_{x}$ are all even integers), we obtain the dispersion of
quasi-particles. The energy of $Z_{2}$ vortex is given by
\begin{equation}
\varepsilon _{k_{1}}=\sqrt{(\xi _{k_{1}}+2g)^{2}-\xi _{k_{1}}^{2}}
\end{equation}%
where
\begin{equation}
\xi _{k_{1}}=2h^{x}[\cos (k_{x}+k_{y})+\cos (k_{x}-k_{y})].
\end{equation}
Here $k_{x}$ and $k_{y}$\ are the wave vectors (The lattice constant has
been set to be unit). The energy gaps of $Z_{2}$ vortex is obtained as $2g%
\sqrt{1-\frac{4h^{x}}{g}}.$ On the other hand, the energies of fermion is
given by
\begin{eqnarray}
\varepsilon _{k_{3}} &=&\sqrt{(\xi _{k_{1}}+4g)^{2}+4(h^{y}\cos k_{x})^{2}}
\\
\varepsilon _{k_{3}} &=&\sqrt{(\xi _{k_{4}}+4g)^{2}+4(h^{y}\cos k_{y})^{2}}
\notag
\end{eqnarray}%
The energy gaps of fermion is obtained as $4g\sqrt{1-\frac{2h^{y}}{g}}$.
Thus one may manipulate the dispersion of $Z_{2}$ vortex and fermion by
tuning the external field.

\section{Effective pseudo-spin model of surface codes}

It is known that the degenerate ground states of $Z_{2}$ topological orders
have identically energy in thermodynamic limit. However, in a finite system,$%
\ $the degeneracy of the ground states is (partially) removed due to
tunneling processes, of which a virtual quasi-particle moves around the
holes before annihilated with the other\cite{k1,wen,ioffe}. In general
cases, one will get very large energy gaps $m_{\alpha }$ ($\alpha =1,2,3,4$)
for all quasi-particles and very tiny energy splitting of the degenerate
ground states $\Delta E$. That is $\Delta E\ll m_{\alpha }$. Based on this
condition ($\Delta E\ll m_{\alpha }$), we may ignore excited states with $%
E>m_{\alpha }$ and consider only the topological degenerate ground states.
Thus we get a $2^{n}$ system as the effective pseudo-spin model of the
surface code. In the followings, we will derive this model step by step.

\begin{figure}[tbp]
\includegraphics[clip,width=0.51\textwidth]{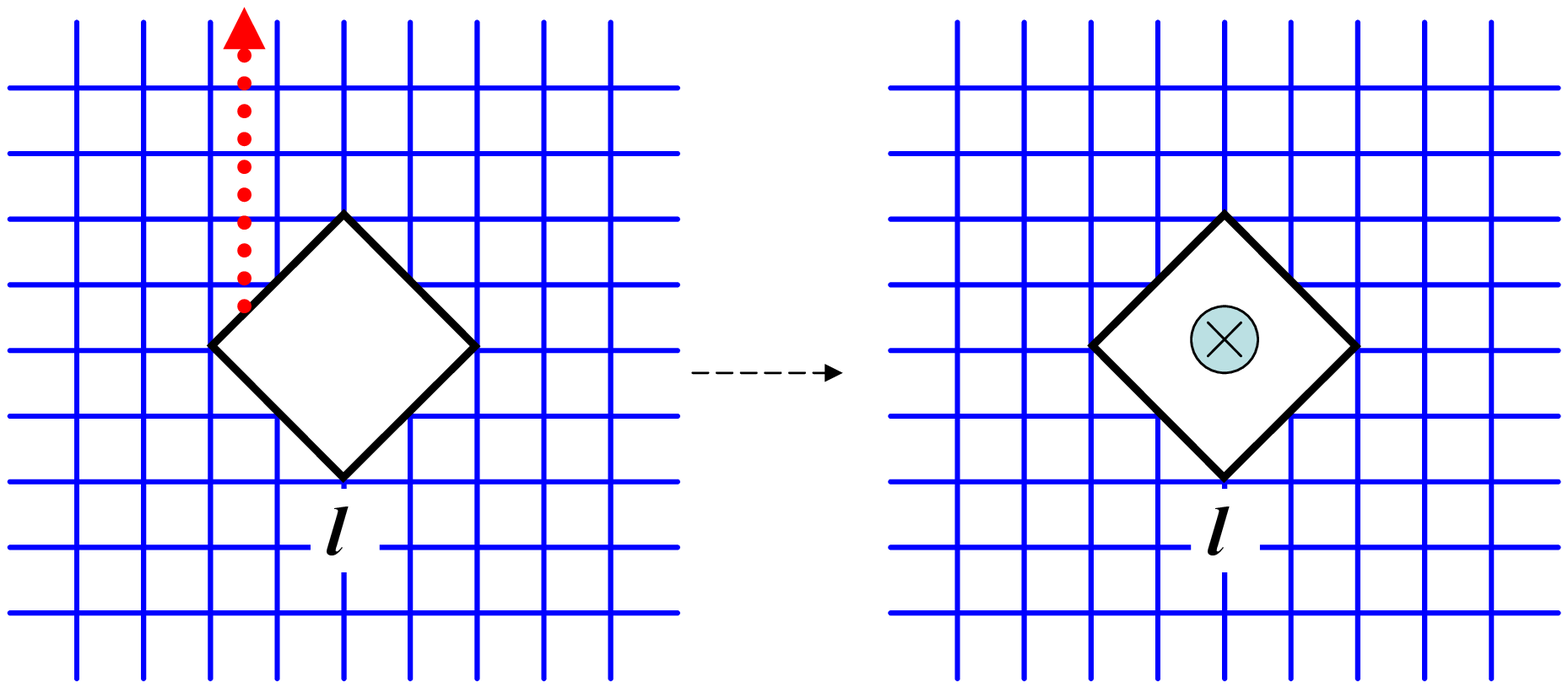}
\caption{Tunneling process of Fermions from one hole to the boundary}
\label{Fig.4}
\end{figure}

In face, the closed string operators $W_{v}(\mathcal{C}_{A})$ and $W_{f}(%
\mathcal{C}_{B})$ can be considered as quantum tunneling processes of
virtual quasi-particle moving along the loops. Let us take the quantum
tunneling process of fermions as an example : at first a pair of the
fermions is created. One fermion propagates around the hole driven by the
operator $\sigma _{i}^{y}$ and then annihilates with the other. Then a
closed string of $\sigma _{i}^{y}$ is left on the tunneling path behind the
virtual fermion, that is just a closed string operator $W_{f}(\mathcal{C}%
_{B})$. Such a process effectively adds the $\pi $ flux to one hole and
changes $m_{l}$ by $1$.

\begin{figure}[tbp]
\includegraphics[clip,width=0.45\textwidth]{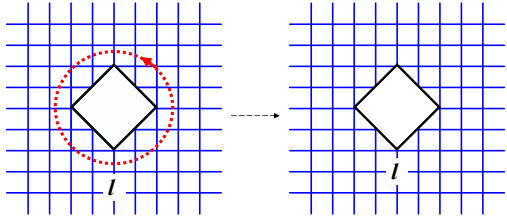}
\caption{Tunneling process of $Z_{2}$ vortex moving around one hole}
\label{Fig.5}
\end{figure}

Firstly we calculate the the effective pseudo-spin model of single qubit
(labeled by $l$-th hole) from tunneling processes. When a virtual fermion
propagates from the boundary of the $l$-th hole to the boundary of the
system, the quantum state $\left(
\begin{array}{c}
\mid \uparrow \rangle _{l} \\
\mid \downarrow \rangle _{l}%
\end{array}%
\right) $ turns into
\begin{equation}
\left(
\begin{array}{c}
\mid \downarrow \rangle _{l} \\
\mid \uparrow \rangle _{l}%
\end{array}%
\right) =\tau _{l}^{x}\left(
\begin{array}{c}
\mid \uparrow \rangle _{l} \\
\mid \downarrow \rangle _{l}%
\end{array}%
\right) .
\end{equation}%
Such process is shown in Fig.4. By a degenerate perturbation approach\cite%
{k1,wen,ioffe,yu}, one may obtain the energy splitting $\Delta E$ of the two
ground states as
\begin{equation}
\delta E=\sum \limits_{j}\text{ }\langle \uparrow \mid _{l}\hat{H}^{\prime }(%
\frac{1}{E_{0}-\hat{H}_{0}}\hat{H}^{\prime })^{j}\mid \downarrow \rangle _{l}
\end{equation}%
which becomes
\begin{equation}
\delta E=2\frac{(h^{y})^{\tilde{L}^{y}}}{(-8g)^{\tilde{L}^{y}-1}}
\end{equation}%
where $\tilde{L}^{y}$ is the length of the shortest path of a fermion from
the boundary of the $l$-th hole to the boundary of the system. See detailed
calculation in Ref.\cite{yu}.

On the other hand, considering a virtual $Z_{2}$ vortex propagating around
the hole (shown in Fig.5), the quantum states $\left(
\begin{array}{c}
\mid \uparrow \rangle _{l} \\
\mid \downarrow \rangle _{l}%
\end{array}%
\right) $ turn into
\begin{equation}
\left(
\begin{array}{c}
\mid \uparrow \rangle _{l} \\
-\mid \downarrow \rangle _{l}%
\end{array}%
\right) =\tau _{l}^{z}\left(
\begin{array}{c}
\mid \uparrow \rangle _{l} \\
\mid \downarrow \rangle _{l}%
\end{array}%
\right) .
\end{equation}%
The corresponding energy difference $\varepsilon $ of the two ground states
is
\begin{equation}
\varepsilon =2\frac{(h^{x})^{\tilde{L}^{x}}}{(-4g)^{\tilde{L}^{x}-1}}
\end{equation}%
where $\tilde{L}^{x}$ is the length of the shortest path of a $Z_{2}$ vortex
around the hole.

Thus the dynamics of such a two-level quantum system (a single qubit) can be
described by a simple effective pseudo-spin Hamiltonian
\begin{eqnarray}
\mathcal{H}_{\mathrm{eff}} &=&\frac{\delta E}{2}(\mid \uparrow \rangle
_{l}\langle \downarrow \mid _{l}+\mid \downarrow \rangle _{l}\langle
\uparrow \mid _{l})+\frac{\varepsilon }{2}(\mid \uparrow \rangle _{l}\langle
\uparrow \mid _{l}-\mid \downarrow \rangle _{l}\langle \downarrow \mid _{l})
\notag \\
&=&\tilde{h}_{l}^{x}\tau _{l}^{x}+\tilde{h}_{l}^{z}\tau _{l}^{z}
\end{eqnarray}%
with $\tilde{h}_{l}^{x}=\frac{\delta E}{2}$ and $\tilde{h}_{l}^{z}=\frac{%
\varepsilon }{2}$.

Secondly we calculate the effective exchange interaction between two qubits.
For simplicity, we consider only the perturbation as $\hat{H}^{\prime
}=h^{x}\sum \limits_{i}\sigma _{i}^{x}+h^{y}\sum \limits_{i}\sigma _{i}^{y}.$
Then from Eq.(13), there exist two different tunneling processes : virtual
fermion propagating from the boundary of $l$-th hole to the boundary of $%
(l+1)$-th hole, virtual $Z_{2}$ vortex propagating around the two holes,
respectively. See Fig.6 and Fig.7. Let us calculate the ground state energy
splitting\ from degenerate perturbation approach.

\begin{figure}[tbp]
\includegraphics[clip,width=0.5\textwidth]{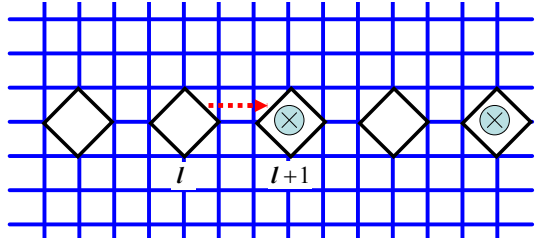}
\caption{Tunneling process of fermions from one hole to another}
\label{Fig.6}
\end{figure}

When a virtual fermion propagating from the boundary of $l$-th hole to the
boundary of $(l+1)$-th hole, the quantum states turn into
\begin{widetext}
\begin{equation*}
\left(
\begin{array}{c}
\mid \uparrow \rangle _{l} \\
\mid \downarrow \rangle _{l}%
\end{array}%
\right) \otimes \left(
\begin{array}{c}
\mid \uparrow \rangle _{l+1} \\
\mid \downarrow \rangle _{l+1}%
\end{array}%
\right) \rightarrow \left(
\begin{array}{c}
\mid \downarrow \rangle _{l} \\
\mid \uparrow \rangle _{l}%
\end{array}%
\right) \otimes \left(
\begin{array}{c}
\mid \downarrow \rangle _{l+1} \\
\mid \uparrow \rangle _{l+1}%
\end{array}%
\right) =\tau _{l}^{x}\mathbf{\otimes }\tau _{l+1}^{x}\left(
\begin{array}{c}
\mid \uparrow \rangle _{l} \\
\mid \downarrow \rangle _{l}%
\end{array}%
\right) \otimes \left(
\begin{array}{c}
\mid \uparrow \rangle _{l+1} \\
\mid \downarrow \rangle _{l+1}%
\end{array}%
\right) .
\end{equation*}%
\end{widetext}Therefore, by the mapping, the pseudo-spin operator of the
tunneling process of fermion corresponds to $\tau _{l}^{x}\mathbf{\otimes }%
\tau _{l+1}^{x}$. See Fig.6. The energy splitting $\delta E$ of the ground
states is obtained as
\begin{equation}
\delta E=J_{l,l+1}^{xx}=\frac{(h^{y})^{L_{yy}}}{(-8g)^{L_{yy}-1}}
\end{equation}%
where $L_{yy}$ is the length of the shortest path from the boundary of $l$%
-th hole to the boundary of $(l+1)$-th hole.

\begin{figure}[tbp]
\includegraphics[clip,width=0.5\textwidth]{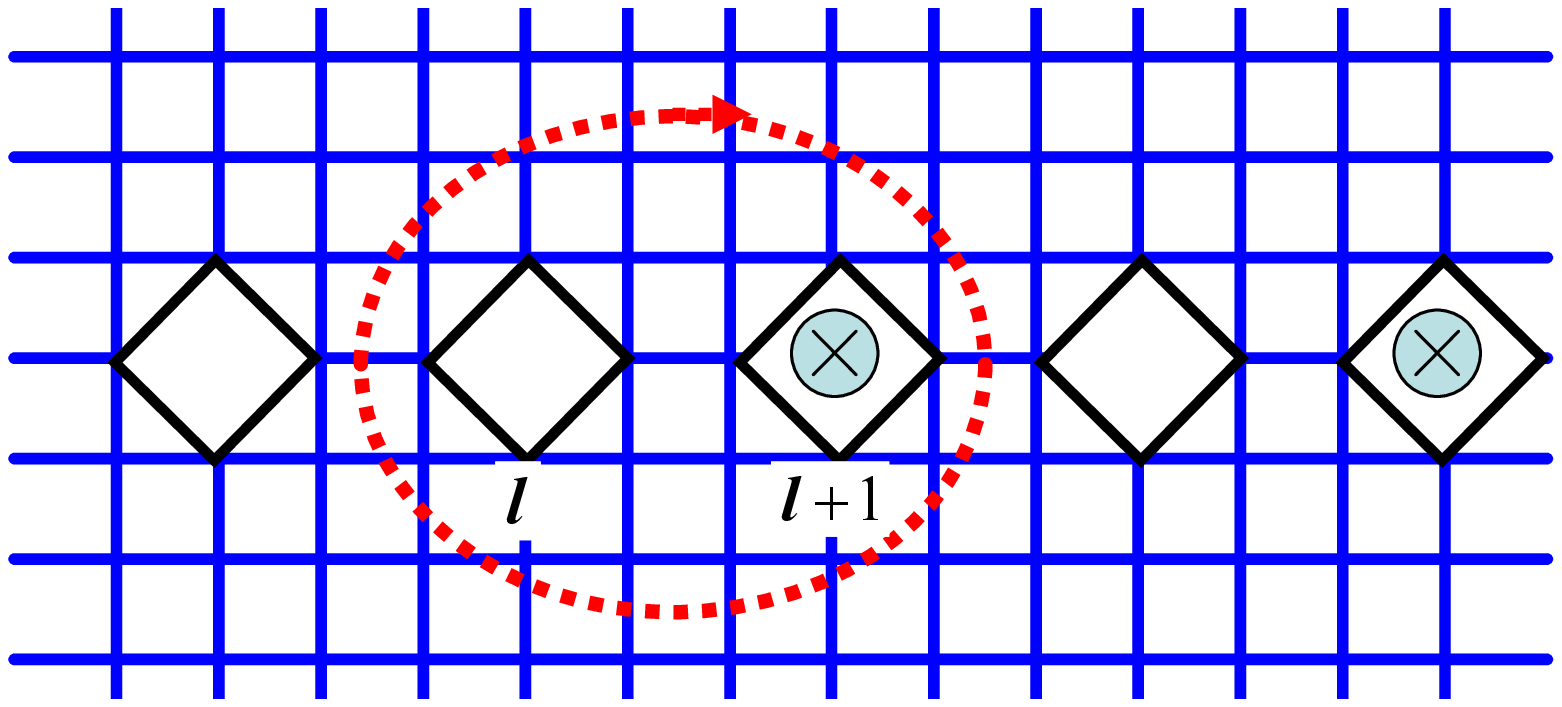}
\caption{Tunneling process of $Z_{2}$ vortex moving around two holes}
\label{Fig.7}
\end{figure}

Similarly, when a virtual $Z_{2}$-vortex propagates around $l$-th and $(l+1)$%
-th holes, the quantum state turns into
\begin{equation*}
\left(
\begin{array}{c}
\mid \uparrow \rangle _{l} \\
\mid \downarrow \rangle _{l}%
\end{array}%
\right) \otimes \left(
\begin{array}{c}
\mid \uparrow \rangle _{l+1} \\
\mid \downarrow \rangle _{l+1}%
\end{array}%
\right) \rightarrow \left(
\begin{array}{c}
\mid \uparrow \rangle _{l} \\
-\mid \downarrow \rangle _{l}%
\end{array}%
\right) \otimes \left(
\begin{array}{c}
\mid \uparrow \rangle _{l+1} \\
-\mid \downarrow \rangle _{l+1}%
\end{array}%
\right) .
\end{equation*}%
The pseudo-spin operators of the tunneling process of $Z_{2}$ vortex
correspond to $\tau _{l}^{z}\mathbf{\otimes }\tau _{l+1}^{z}$. The tunneling
amplitude is
\begin{equation}
J_{l,l+1}^{zz}=\frac{(h^{x})^{L_{xx}}}{(-4g)^{L_{xx}-1}}
\end{equation}%
where $L_{xx}$ is the length of the shortest path round both $l$-th and $%
(l+1)$-th holes. Thus under the perturbation $\hat{H}^{\prime
}=h^{x}\sum \limits_{i}\sigma _{i}^{x}+h^{y}\sum \limits_{i}\sigma _{i}^{y},$
the total effective Hamiltonian of the exchange interaction becomes

\begin{equation}
\mathcal{H}_{\mathrm{eff}}\simeq \sum \limits_{l}(J_{l,l+1}^{xx}\tau
_{l}^{x}\tau _{l+1}^{x}+J_{l,l+1}^{zz}\tau _{l}^{z}\tau _{l+1}^{z}).
\end{equation}
\

Finally, for a chain of ${n}${-}hole{, the degenerate ground states can be
mapped onto a model of a }${n}${-pseudo-spin chain. }By ignoring the next
nearest neighbor coupling terms, the effective model of the Kitaev
Toric-code model\ under the perturbation $\hat{H}^{\prime }=h^{x}\sum
\limits_{i}\sigma _{i}^{x}+h^{y}\sum \limits_{i}\sigma _{i}^{y}$ is
naturally an anisotropy Heisenberg model
\begin{equation}
\mathcal{H}_{\mathrm{eff}}\simeq \sum \limits_{l}(J_{l,l+1}^{xx}\tau
_{l}^{x}\tau _{l+1}^{x}+J_{l,l+1}^{zz}\tau _{l}^{z}\tau _{l+1}^{z})+\sum
\limits_{l}(\tilde{h}_{l}^{x}\tau _{l}^{x}+\tilde{h}_{l}^{z}\tau _{l}^{z}).
\label{eff}
\end{equation}%
The effective Hamiltonian in Eq.(\ref{eff}) indicates that the Kitaev
Toric-code model is an example of so-called \emph{topological order with
controllable dispersion of quasi-particles}. For example, if one adds the
external field along $y$-direction only encircling two holes, $\hat{H}%
^{\prime }=h^{y}\sum \limits_{i}\sigma _{i}^{y}$, the effective model is
reduced into
\begin{equation}
\mathcal{H}_{\mathrm{eff}}\simeq J_{1,2}^{xx}\tau _{1}^{x}\tau _{2}^{x}+%
\tilde{h}_{1}^{x}\tau _{1}^{x}+\tilde{h}_{2}^{x}\tau _{2}^{x}.
\end{equation}%
So one may adjust each parameters in the effective Hamiltonian by
controlling the local distribution of the external field along special
direction.

\section{Topological quantum computation with surface codes}

To design a topological quantum computer, one needs to do arbitrary unitary
operations on the surface codes. Then by adding the specific perturbations
to the Kitaev Toric-code model, $H^{\prime },$ one can change different
quasi-particles' hopping and then manipulate the surface codes by
controlling tunneling splitting of degenerate ground states. In this part we
show the initialization, the unitary transformation and the measurement.

\subsection{Initialization}

Firstly we will show how to initialize the system into the quantum state $%
\mid \uparrow ,...,\uparrow \rangle $. The basic idea is \emph{to polarize
all pseudo-spins by adding an effective field along x-direction and then
removal it slowly}. This process will occur according to the Hamiltonian
\begin{equation}
H^{\prime }{=h}^{x}{(t)\sum \limits_{i}}\mathbf{\sigma }_{i}^{x}
\end{equation}%
where ${h}^{x}{(t)=h}_{0}{e}^{-t_{0}/t}$. At the beginning, there is a
finite external field ${h}^{x}{(t)\rightarrow h}_{0}$, $t\rightarrow -\infty
.$ At the time $t=0,$ the external field disappears, ${h^{x}(t)=0}$. From Eq.%
\ref{eff}, we get the effective pseudo-spin Hamiltonian of the surface codes
\begin{equation}
\mathcal{H}_{\mathrm{eff}}\simeq \sum \limits_{l}J_{l,l+1}^{zz}\tau
_{l}^{z}\tau _{l+1}^{z}+\sum \limits_{l}\tilde{h}_{l}^{z}(t)\tau _{l}^{z}
\end{equation}%
where $J_{l,l+1}^{zz}=\frac{[h^{x}(t)]^{L_{xx}}}{(-4g)^{L_{xx}-1}}$ and $%
\tilde{h}_{l}^{z}(t)=\frac{[h^{x}(t)]^{\tilde{L}^{x}}}{(-4g)^{\tilde{L}%
^{x}-1}}.$ Here ${h}_{0}$ is positive and $\tilde{L}^{x}$ is an even number.
Then if the system evolves \emph{adiabatically and continuously} from high
temperature to the ground state, after a long time, the final state can be a
pure state of the topological order $\mid \uparrow ,...,\uparrow \rangle $
which becomes the initial state prepared for TQC.

\subsection{Unitary operations}

Secondly we discuss how to do an arbitrary unitary transformation on the
surface code\cite{du1,zhang,zoller}. The key point here is that \emph{the
unitary operations can be achieved by controlling the external field along
particular direction within fixed times.}

A general pseudo-spin rotation operator of $l$-th qubit is defined by
\begin{equation}
U_{l}(\theta ,\text{ }\varphi ,\text{ }\gamma )=e^{-\frac{i}{\hbar }\gamma
\tau _{l}^{z}}e^{-\frac{i}{\hbar }\varphi \tau _{l}^{x}}e^{-\frac{i}{\hbar }%
\theta \tau _{l}^{z}}
\end{equation}%
where $\gamma =\tilde{h}_{l}^{z}\Delta {t}_{\gamma },$ $\theta =\tilde{h}%
_{l}^{z}\Delta {t}_{\theta }$ and $\varphi =\tilde{h}_{l}^{x}\Delta {t}%
_{\varphi }.$ One can use external field along different directions
encircling only $l$-th hole to do TQC : firstly applying the external field
along \textrm{y}-direction at an interval $\Delta {t}_{\theta }{.}$ The
effective Hamiltonian becomes $\mathcal{H}_{\mathrm{eff}}=\tilde{h}%
_{l}^{z}\tau _{l}^{z}$. Then, we swerve the external field along \textrm{x}%
-direction at an interval $\Delta {t}_{\varphi }{.}$ The effective
Hamiltonian becomes $\mathcal{H}_{\mathrm{eff}}=\tilde{h}_{l}^{x}\tau
_{l}^{x}$. Finally, the external field along \textrm{y}-direction is added
at an interval $\Delta {t}_{\gamma }{.}$ The effective Hamiltonian becomes $%
\mathcal{H}_{\mathrm{eff}}=\tilde{h}_{l}^{z}\tau _{l}^{z}$.

Using such method, one can reach certain quantum operations demanded by TQC
and have the ability to carry out gate operations onto the 'local' qubit ($l$%
-th hole) at will,
\begin{equation}
\mid \Psi \rangle _{l}=\alpha \mid \uparrow \rangle _{l}+\beta e^{i\phi
}\mid \downarrow \rangle _{l}
\end{equation}
with $\alpha ,$ $\beta \geq 0$ ($\alpha ^{2}+\beta ^{2}=1$). An example is
the $\frac{\pi }{8}$ gate of a local qubit ($l$-th hole), of which we have
the general pseudo-spin rotation operator as%
\begin{equation*}
U_{l}(\gamma {=\frac{\pi }{8},}\text{ }\theta {=0,}\text{ }\varphi =\frac{%
\pi }{8}).
\end{equation*}%
One can firstly applying a global external field along \textrm{y}-direction
at an interval $\Delta {t}_{\varphi }=\frac{\pi }{8\tilde{h}_{l}^{x}}{.}$
Then, we swerve the global external field along \textrm{x}-direction at an
interval $\Delta {t}_{\gamma }=\frac{\pi }{8\tilde{h}_{l}^{z}}{.}$
Similarly, one may design the global Hadamard gate as a special pseudo-spin
rotation operator on each qubit,
\begin{equation*}
U_{l}(\gamma {=\frac{\pi }{4},}\text{ }\theta {=\frac{7\pi }{4},}\text{ }%
\varphi =\frac{\pi }{4}).
\end{equation*}

Thus, {in principle, people are capable of to do arbitrary unitary
transformation on the protected subspace by controlling the external field
on given regions (for example, a close loop around one or more hole). }

\subsection{Measurement}

Thirdly we discuss the measurement of an arbitrary quantum states of the
surface codes. The central point is \emph{to measure the expected values of
pseudo-spin operators by observing the quasi-particles' interferences from
Aharonov--Bohm (AB) effect.}

To determine $\alpha ,$ $\beta $ and $\phi $ of quantum state of the $l$%
-hole, $\mid \Psi \rangle _{l}=\alpha \mid \uparrow \rangle _{l}+\beta
e^{i\phi }\mid \downarrow \rangle _{l},$ we need to observe both fermion
interference and $Z_{2}$ vortex interference. Fig.8 is a scheme to show the
AB interference.

Firstly we detect the value of $\left \langle \tau _{l}^{z}\right \rangle
=\langle \Psi \mid _{l}\tau _{l}^{z}\mid \Psi \rangle _{l}$ to determine $%
\alpha $ and $\beta $ by AB effect from $Z_{2}$ vortex-interference. To
observe the AB interference, we add a small external field, $%
h^{x}\rightarrow 0$ and $h^{y}=0$. Now $Z_{2}$ vortex begin to hop. There
exist symmetrical paths from both sides of the hole. For example, $\gamma
_{1}$ and $\gamma _{2}$ shown in Fig.8 are two symmetrical paths. Then the
symmetrical trajectories will contribute to the transition amplitude $%
T_{i,j} $ according to :
\begin{equation}
T_{i,j}={\left \vert \psi _{i,j}^{\gamma _{1}}\right \vert ^{2}}+{\left
\vert \psi _{i,j}^{\gamma _{2}}\right \vert ^{2}}+2\epsilon \left \vert \psi
_{i,j}^{\gamma _{2}}\psi _{i,j}^{\gamma _{1}}\right \vert
\end{equation}%
where $\psi _{i,j}^{\gamma _{2}}$ and $\psi _{i,j}^{\gamma _{1}}$ are the
wave functions of $Z_{2}$ vortex of the two trajectories. For the ground
state $\mid \uparrow \rangle _{l}$, $\epsilon $ is unit. However, for the
ground state with $\pi $-flux inside the hole $\mid \downarrow \rangle _{l},$
we have $\epsilon =-1.$ Then we can distinguish these two cases. For two
symmetrical paths $\psi _{i,j}^{\gamma _{2}}=\psi _{i,j}^{\gamma _{1}}{=t}%
_{f}$, we get a probability $\alpha ^{2}$ for $\mid \uparrow \rangle _{l}$
with $T_{i,j}=4{t_{f}^{2}}$ and a probability $\beta ^{2}$ for $\mid
\downarrow \rangle _{l}$ with $T_{i,j}=0$.

\begin{figure}[tbp]
\includegraphics[clip,width=0.35\textwidth]{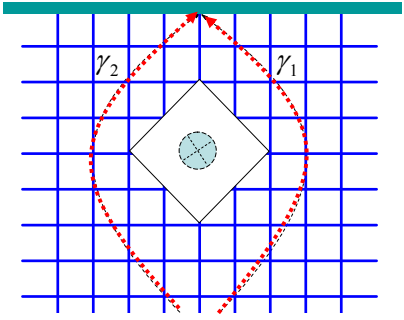}
\caption{The scheme of the interference around $l$-th hole. }
\label{Fig.8}
\end{figure}

On the other hand, one can detect the value of $\left \langle \tau
_{l}^{x}\right \rangle =\langle \Psi \mid _{l}\tau _{l}^{x}\mid \Psi \rangle
_{l}$ to determine the parameter $\phi $ by observing fermion interference.
To observe the AB interference of fermion, we add a small external field, $%
h^{y}\rightarrow 0$. The wave function of fermion has a periodic boundary
condition from the hole to the boundary of the system for the ground state $%
\mid \uparrow ^{\prime }\rangle _{l}=\frac{1}{\sqrt{2}}\mid \uparrow \rangle
_{l}+\mid \downarrow \rangle _{l}$ and an anti-periodic boundary condition
for the ground state $\mid \downarrow ^{\prime }\rangle _{l}=\frac{1}{\sqrt{2%
}}\mid \uparrow \rangle _{l}-\mid \downarrow \rangle _{l}.$ Then an
arbitrary state $\alpha \mid \uparrow \rangle _{l}+\beta e^{i\phi }\mid
\downarrow \rangle _{l}$ is re-written into
\begin{equation}
\sqrt{\frac{1}{2}+\alpha \beta \cos \phi }e^{i\phi ^{\prime }}\mid \uparrow
^{\prime }\rangle _{l}+\sqrt{\frac{1}{2}-\alpha \beta \cos \phi }\beta
e^{i\phi ^{\prime \prime }}\mid \downarrow ^{\prime }\rangle _{l}
\end{equation}%
where
\begin{equation}
\phi ^{\prime }=\arctan (\frac{\sin \phi }{\beta \cos \phi +\alpha })
\end{equation}%
and
\begin{equation}
\phi ^{\prime \prime }=\arctan (\frac{\sin \phi }{\beta \cos \phi -\alpha }).
\end{equation}%
For two symmetrical paths (one from the hole, the other not), we get a
probability $(\frac{1}{2}+\alpha \beta \cos \phi )$ for $\mid \uparrow
^{\prime }\rangle _{l}$ with $T_{i,j}=4{t_{v}^{2}}$ and a probability $\frac{%
1}{2}-\alpha \beta \cos \phi $ for $\mid \downarrow ^{\prime }\rangle _{l}$
with $T_{i,j}=0$. As a result, we determine the parameters $\alpha ,$ $\beta
$ and $\phi $ of an arbitrary state $\mid \mathrm{vac}\rangle =\alpha \mid
\uparrow \rangle _{l}+\beta e^{i\phi }\mid \downarrow \rangle _{l}$.

\begin{figure}[tbp]
\includegraphics[clip,width=0.48\textwidth]{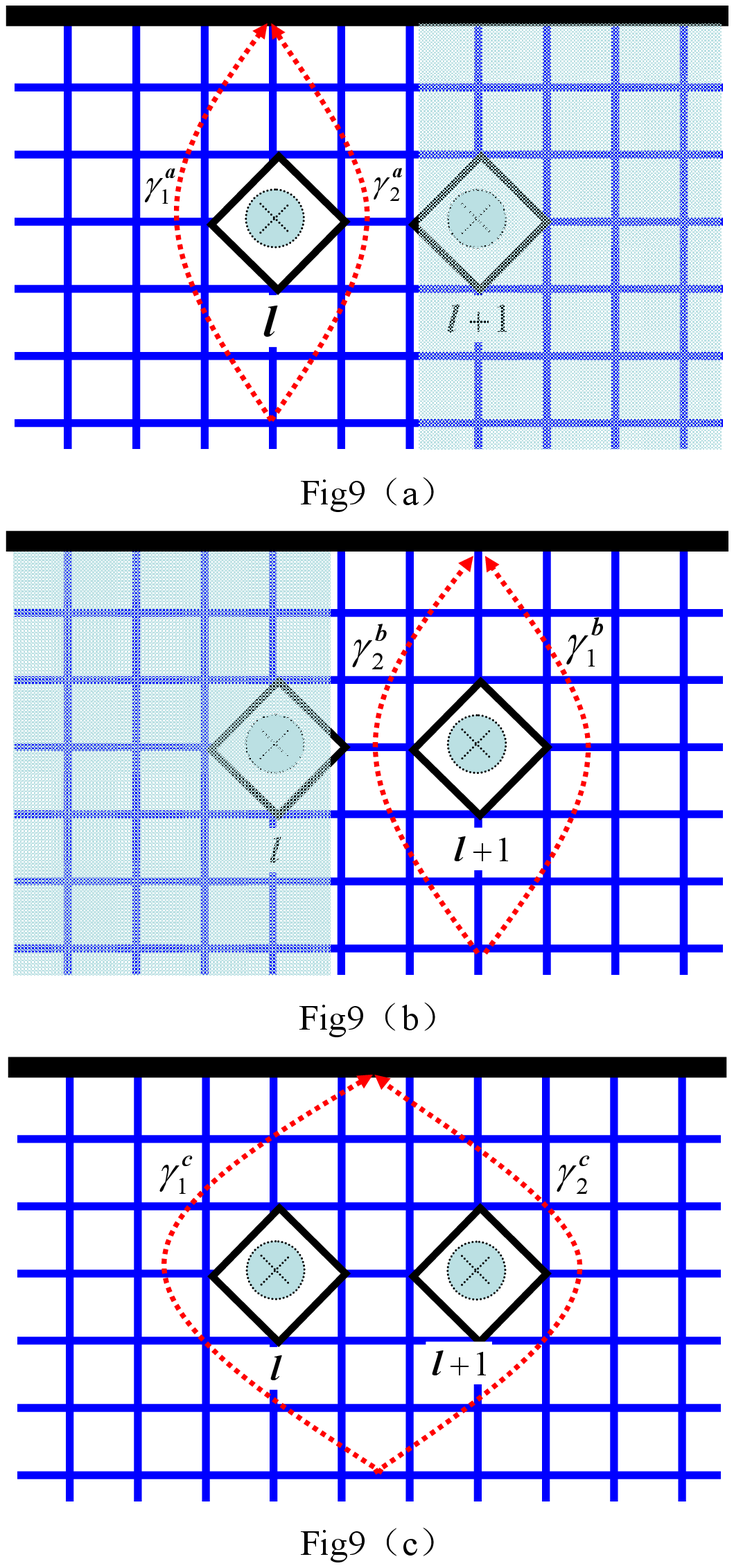}
\caption{The scheme of the interference around $l$-th hole in (a), the
interference around $l+1$-th hole in (b), and the interference around two
holes in (c). Z2 vortex cannot move in shadow region. }
\label{Fig.9}
\end{figure}

The situation becomes more complex for $2$-qubit. For $2$-qubit with $2$
holes ($l$-th and $l+1$-th), a general entangled quantum state is given by
\begin{eqnarray*}
\alpha _{1} &\mid &\uparrow \rangle _{l+1}\otimes \mid \uparrow \rangle
_{l}+\alpha _{2}e^{i\phi _{2}}\mid \uparrow \rangle _{l+1}\otimes \mid
\uparrow \rangle _{l} \\
+\alpha _{3}e^{i\phi _{3}} &\mid &\uparrow \rangle _{l+1}\otimes \mid
\downarrow \rangle _{l}+\alpha _{4}e^{i\phi _{4}}\mid \downarrow \rangle
_{l+1}\otimes \mid \uparrow \rangle _{l}
\end{eqnarray*}%
Here $\alpha _{i}$ and $\phi _{i}$ are all real number. Because of the
constraint, $\sum_{i}\alpha _{i}^{2}=1,$ there are totally $2(2^{2}-1)=6$
independent parameters. One can detect the value of $\left \langle \tau
_{l}^{z}\right \rangle ,$ $\left \langle \tau _{l+1}^{z}\right \rangle ,$ $%
\left \langle \tau _{l}^{z}\tau _{l+1}^{z}\right \rangle $ to determine $%
\alpha _{i}$ $(i=1,2,3,4)$ from the AB effect of $Z_{2}$-vortex-interference
by $2^{2}-1=3$ times measurements. As shown in Fig.9, there are $3$ cases of
the $Z_{2}$ vortex-interference of different paths : $\gamma _{1}^{a}$ and $%
\gamma _{2}^{a}$ are two symmetrical paths around the hole $l$; $\gamma
_{1}^{b}$ and $\gamma _{2}^{b}$ are two symmetrical paths around the hole $%
l-1$; $\gamma _{1}^{c}$ and $\gamma _{2}^{c}$ are two symmetrical paths
around both hole $l$ and hole $l-1.$ To do the observations, we only add a
small external field in the regions without shadow, $h^{x}\rightarrow 0$ and
$h^{y}=0$ (In the shadow regions, there is no external field, $h^{x}=h^{y}=0$%
)$.$ So the $Z_{2}$-vortex is guided moving around given holes. Similarly,
one can detect the value of $\left \langle \tau _{l}^{x}\right \rangle ,$ $%
\left \langle \tau _{l+1}^{x}\right \rangle ,$ $\left \langle \tau
_{l}^{x}\tau _{l+1}^{x}\right \rangle $ determine $\phi _{i}$ $(i=2,3,4)$
from the AB effect of fermion-interference by $3$ times measurements.

For $n$-qubit with $n$ holes, a general entangled quantum state is given by
\begin{equation}
\sum_{i}\alpha _{i}e^{i\phi _{i}}\mid \uparrow \rangle _{i}\otimes \mid
\uparrow \rangle _{l}\otimes \mid \uparrow \rangle _{l+1}
\end{equation}%
Here $\alpha _{i}$ and $\phi _{i}$ are all real number. Because of the
constraint, $\sum_{i}\alpha _{i}^{2}=1,$ there are totally real $2(2^{n}-1)$
independent parameters by changing the paths of quasi-particles. So people
needs to do $2(2^{n}-1)$ times measurement to determine the entangle state
by changing different paths of AB\ interferences. One can determine $\alpha
_{i}$ from $\frac{n(n+1)}{2}$ times measurements of $Z_{2}$
vortex-interference (to detect $\left \langle \tau _{l}^{z}\right \rangle ,$ $%
\left \langle \tau _{l}^{z}\tau _{l+1}^{z}\right \rangle ,$ $\left \langle \tau
_{l}^{z}\tau _{l+1}^{z}\tau _{l+2}^{z}\right \rangle ,$...$,$ $\left \langle
\prod_{l}\tau _{l}^{z}\right \rangle $ ) and $\phi _{i}$ from $\frac{n(n+1)}{2%
}$ times measurements of fermion-interference (to detect $\left \langle \tau
_{l}^{x}\right \rangle ,$ $\left \langle \tau _{l}^{x}\tau
_{l+1}^{x}\right \rangle $), respectively. So one cannot determine all the
parameters by the quasi-particles interferences if $n>3$. It is still an
unsolved problem to measure a general entangled quantum state of $n$-qubit
with $n$ ( $n>3$) holes.

\subsection{Errors}

Finally we discuss the errors and the constraint on our proposal.

Errors mainly come from the thermal effect. At finite temperature, $Z_{2}$
vortices are excited, their moving around the holes leads to errors, as
causes \emph{"thermal hopping"} from one degenerate ground state to another.
\emph{The} \emph{decoherence time} $t_{\mathrm{de}}$ has been roughly
estimated by the time to stretch a pair of $Z_{2}$ vortices over a distance
equal to the average inter-particle separation (This is because the energy
gap of $Z_{2}$ vortex is smaller than that that of fermion). At low
temperature, one can estimate that $t_{\mathrm{de}}$ is about $t_{0}e^{\frac{%
4g}{k_{B}T}}$ where $t_{0}=\frac{L_{p}}{v^{\ast }}$ is the time scale for $%
Z_{2}$ vortex moving the length of tunneling pathes, $L_{p}$ \cite{ort}. $%
v^{\ast }$ is the average speed of $Z_{2}$ vortex, which is estimated by $%
v^{\ast }\sim \sqrt{\frac{k_{B}T}{M_{\mathrm{eff}}}}$ where $M_{\mathrm{eff}%
}\simeq \left( 2h^{x}\right) ^{-1}$ is the effective mass of $Z_{2}$ vortex.

\begin{figure}[tbp]
\includegraphics[clip,width=0.4\textwidth]{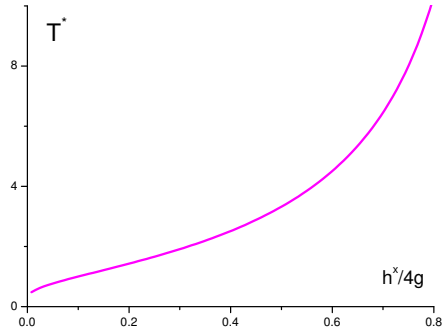}
\caption{The crossover temperature for an example of tunneling process of Z2
vortex. The scale of temperature is $\frac{k_{B}L_{p}}{4g}$}
\label{Fig.12}
\end{figure}

So there must exist a \emph{crossover temperature} $T^{\ast }$ from thermal
hopping to quantum tunneling. Above $T^{\ast }$, the decay rate of the
quantum states is determined by process of thermal activation, which is
governed by the Arrhenius law,
\begin{equation}
t_{0}^{-1}e^{-\frac{4g}{k_{B}T}}.
\end{equation}%
Therefore at high temperature the errors proliferate and one cannot get
reliable TQC. Below $T^{\ast }$, quantum tunneling processes dominate, the
rate of which goes as $e^{-B}$ where $B$ is about
\begin{equation}
B\sim \max (L_{p}\ln \frac{4g}{\left \vert h_{x}\right \vert },\text{ }%
L_{p}\ln \frac{8g}{\left \vert h_{y}\right \vert }).
\end{equation}%
Ignoring the prefactor and equating the exponents, one obtains
\begin{equation}
T^{\ast }=\frac{4g}{k_{B}B}.  \label{t}
\end{equation}%
Fig.10 shows the crossover temperature via $h_{x}$ for a quantum tunneling
process of $Z_{2}$ vortex. Thus if the temperature is kept far below $%
T^{\ast }$, $T\ll T^{\ast }$, one may do unitary operations safely.

However, because of the errors from the stochastic fields, we still get in
trouble on storing quantum information by the surface codes. To store
quantum information, all quantum tunneling processes need to be suppressed
as low as possible, $e^{-B}\rightarrow 0$. That means the external fields
should be removed, $h^{x}\equiv h^{y}\equiv 0$. Now the estimation of the
crossover temperature in Eq.\ref{t} is invalid. Without external fields, due
to the diverge effective masses of $Z_{2}$ vortex, $M_{\mathrm{eff}%
}\rightarrow \infty ,$ one may store the quantum information for arbitrary
long time, $t_{\mathrm{de}}\rightarrow \infty $. Whereas, the stochastic
noise fields leads to a finite decoherence time $t_{\mathrm{de}}$. Thus to
get a long-lived quantum information, both the temperature and the
stochastic noise fields should be suppressed below a threshold. See detail
in Ref.\cite{sta}.

\section{Summary}

In this paper, we find an alternative way towards designing a quantum
computer{\ that may be possible to incorporate intrinsic fault tolerance.
Using the Kitaev toric-code model as an example, we obtain the effective
pseudo-spin model that can be mapped onto the anisotropic Heisenberg model
of a pseudo-spin chain in external field. Then one may tune the parameters
of the effective pseudo-spin model by controlling tunneling processes of the
surface codes by applying external field along special direction on
lattices. In particular, }five criterions to build a quantum computer are
satisfied :

\begin{enumerate}
\item Scalability of extendible qubits : {The qubits is }the so-called
surface codes (two degenerate ground states of $Z_{2}$ topological orders on
a plane with a hole). So the quantum computer is just a line of holes in the
$Z_{2}$ topological order.

\item Initialization (Creation of highly entangled states) : We may polarize
the pseudo-spins by adding an effective field along x-direction and then
removing it slowly.

\item Local operations on multi-qubit : The unitary operations can do by
controlling the external field along particular direction within fixed times.

\item Measurement of entangled states : We may measure the expected values
of pseudo-spin operators by observing the quasi-particles' interferences
from AB effect.

\item Low decoherence : Below the crossover temperature $T^{\ast },$\ the
decoherence processes will be controlled as low as possible.
\end{enumerate}

Finally we discuss the realization of the Kitaev toric-code model. Because
it can be regarded as an effective model of the Kitaev model on a two
dimensional hexagonal lattice, one may realize the Kitaev model firstly. The
realization of the Kitaev model has been proposed in an optical lattice of
cold atoms in Ref.\cite{du,zo1} and in Josephson junction array of a
superconductor in Ref.\cite{you}. So it is possible to design a quantum
computer in these system in the future.

In general. to design a topological quantum computer via quantum tunneling
effect, there are four necessary conditions : 1) $Z_{2}$ topological order
as the ground states; 2) controllable dispersion of quasi-particles; 3)
space with nontrivial topological structure (manifold with high genus or
multi-hole); 4) low temperature and low noise. Thus considering these
conditions, one may do TQC based on other models.

\begin{acknowledgments}
This research is supported by NCET, NFSC Grant no. 10874017. The
author would like to thank D. L. Zhou, B. Zeng, Y. Shi for helpful
conversations.
\end{acknowledgments}


\begin{thebibliography}{99}
\bibitem{edge} There are two types of edge states of the Kitaev toric-code
model. For a boundary parallels x-direction or y-direction, there exist the
gapless edge states described by Majorana fermions\cite{wen5,kou2}. For a
boundary along diagonal directions ($\hat{e}_{x}-\hat{e}_{y}$ or $\hat{e}%
_{x}+\hat{e}_{y}$), there is no stable edge state. As a result, in this
paper the boundary of both the holes and the system belong to belongs to the
second type (along diagonal directions).
\end{thebibliography}
\end{document}